\documentstyle[12pt]{article}
\begin{document}
\large

\begin{center}
{\Large \bf   Differential constraints and exact solutions \\
 of nonlinear diffusion equations}
\end{center}

\begin{center}
{\bf Oleg V. Kaptsov and Igor V. Verevkin}

 Institute of Computing Modeling RAS, Academgorodok,\\
  660036, Krasnoyarsk, Russia

   E-Mail: kaptsov@ksc.krasn.ru
\end{center}

{\bf Abstract}

The differential constraints are applied to obtain explicit
solutions of nonlinear diffusion equations. Certain linear
determining equations with parameters are used to find such
differential constraints. They generalize the determining
equations used in the search for classical Lie symmetries.

\vspace{05mm}


\vspace{15mm}
 {\bf 1. Introduction.}

 Differential constraints arisen originally in the
theory of partial differential equations of the first order. In
particular Jacobi used differential constraints to find the total
integral of nonlinear equation $$
F(x_1,...,x_n,z,z_{x_1},...,z_{x_n}) = 0 ,  $$
 K\"{o}nig applied them
to the equation of the second order \cite{gou1}. They required
that the corresponding over-determined system was compatible. The
general theory of over-determined systems was developed by
Delassus, Riquier, Cartan, Ritt, Kuranishi, Spencer and others.
One can find references in the book of Pommaret  \cite{pom2}. Now
the applications of over-determined systems include such diverse
fields as differential geometry, continuum mechanics and nonlinear
optics.

 The general formulation of the method of differential constraints requires that
 the original system of partial differential equations
$$ F^1 = 0,..., \quad F^m = 0      \eqno (1) $$
 be enlarged by
appending additional differential equations (differential
constraints)
 $$ h_1 = 0,..., \quad h_p = 0,  \eqno (2) $$
 such that the over-determined system (1), (2) satisfies some conditions
of compatibility.

One can derive many exact solutions of partial differential
equation by means of differential constraints. It was particularly
shown in \cite{and3} that some soliton solutions can be found
using differential constraints. Olver and Rosenau \cite{olv4},
Olver \cite{olv5}, Kaptsov \cite{and3}, Levi and Winternitz
\cite{lev6} show that many reduction methods such as non-classical
symmetry groups, partial invariance, separation of variables, the
Clarkson--Kruskal direct method can be included into the method of
differential constraints. In practice, methods based on the
Riquier--Ritt theory of over-determined systems of partial
differential equations may be difficult. The problem of finding
all differential constraints compatible with certain equations can
be more complicated than the investigation of the original
equations.

     It was recently proposed a new method for finding differential constraints which uses
linear determining equations. These equations are more general
than the classical determining equations for Lie generators
\cite{ovs7} and depend on some parameters.
 Given an evolution equation
$$ u_t = F(t,x,u,u_1,\dots,u_n) ,   \eqno (3) $$
 where $u_k = \frac{\partial^k u}{\partial x^k}$,  then according
to \cite{kap8} the linear determining equation corresponding to
(3) is of the form
 $$D_t(h) = \sum^{N}_{i=0} \sum^{i}_{k=0} b_{ik}
D^{i-k}_x(F_{u_{N-k}}) D^{N-i}_x(h) ,       b_{ik}\in R.\eqno(4)$$
 Here and throughout $D_t, D_x$ are the operators of
total differentiation with respect to $t$ and $x$. Equality (4)
must hold for all solutions of (3). The function $h$ may depends
on $t, x, u, u_1,\dots,u_p$. The number $p$ is called the order of
the solution of equation (4).
 If we have some solution $h$ then corresponding differential constraint is
$$ h =0.   \eqno(5) $$
 It was also shown in \cite{kap8} that equations
(4) and (5) constitute the compatible system. Thus we sketch the
derivation of some solutions to the evolution equation (4).\\
 (I) Find solutions of the linear determining equations
(4).\\
 (II) Fixing the function $h$, we obtain differential
constrain (5).\\
 (III) Find the general solution of (5) which
includes some arbitrary functions $a_i$ depending on $t$.\\
(IV)Substitute the general solution into (4). It leads to ordinary
differential equations for functions $a_i$.\\
 (V) Solve the ordinary differential equations and obtain a solution of the evolution
equation (4).

 In this paper we start with determination of the
solutions of linear determining equations of the second and third
orders for the nonlinear diffusion equation
 $$u_t = (u^k u_x)_x + f(u).  \eqno(6) $$
 These solutions exist only if $f$ belongs to the special forms. Then we
use the obtained functions $h$ to find solutions of equation (6).
In final section we derive exact solutions of two-dimension
equation $$ u_t = \Delta \ln (u) .$$

\vspace{15mm}
 {\bf 2. Solutions of linear determining equations.}

The nonlinear diffusion equation
 $$u_t = (Q(u)u_x)_x + f(u). \eqno (7)  $$
 often arises in the description of various physical processes. The
group classification of the equation has been carried out in
\cite{dor9}. Some exact solutions of (7) can be found in
\cite{gal10,cla11}. In physical applications $Q$ is usually taken
to be a power function. In this section we consider the equation
 $$u_t = (u^q u_x)_x + f(u), \eqno(8) $$
 where $f$ is an differentiable function, $q \neq 0$. If $q=-2$,
$f=u$ or $f=const$ then the equation (8) can be linearized. We
shall not discuss this case here.

The linear determining equation, which corresponds to (8), is
 $$ D_{t}h=u^{q}D_{x}^{2}h+ b_{1} q u_{x} u^{q-1} D_{x}h+ (b_{3} q
u^{q-1} u_{xx}+ b_{2} q (q-1) u^{q-2} u_{x}^{2}+b_{4} f_{u}) h ,
   \eqno(9)$$
 where $b_1,\dots,b_4 \in R.$ We shall seek solutions to (9) in the
form
 $$h=u_{n}+g(t,x,u,\dots,u_{n-1}), $$
 where  $n\geq 2,$ $u_{k}=\frac{ {\partial}^{k}u }{\partial x^{k}}.$
The method for finding solutions is very similar to the standard
procedure applied in the group analysis of differential equations
\cite{olv12} and only one of all possibilities is described here
for the sake of brevity.

We set $n =2.$ First, let us express all $t$-derivatives in (9)
using (8). As a result, the left-hand side of (3.4) becomes a
polynomial with respect to $u_3, u_2$. The polynomial must
identically vanish. Collecting similar terms we obtain the
following relations for the coefficients of $u_3$ and $u_2^2$
 $$q (b_{1}-4)=0, \quad u g_{u_{1}u_{1}}+q(b_{3}-3)=0.$$
 Thus $b_1 = 4$ and $g$ can be represented as follows
 $$g=\frac{(3-b_{3})q}{2u} u_{1}^{2}+a(u,t,x)u_{1}+g_{1}(u,t,x);$$
here $a$ and $g_{1}$  must be functions of $u, t$ and  $x$ alone.
Collecting the coefficients of $u_{2}u_{1}^{2}$ and $u_{2}u_{1}$,
we have the equations
 $$ 2b_{2}q - 2b_{2} - b_3^2 q + b_{3}q + 4b_{3} - 6q =0, \eqno(10)$$
$$ 2ua_{u} + q(b_{3} + 1)a = 0. $$
 In follows from the last equation that
 $$a=a_{1}(t,x)u^{-\frac{(1+b_3)}{2}q},$$
where $a_1$ is a function of $t$ and $x$.
 Next we consider the coefficient $u_1^3$ and obtain equation
$$4b_{2}q-4b_{2}+b_{3}^{2}q-4b_{3}q+2b_{3}-9q+6=0. \eqno (11)$$
From (10) and (11) it follows that $b_3=1$ or $b_3=\frac{q+2}{q}.$

Assuming $b_3=1$, we obtain $b_{2}=\frac{3q-2}{q-1}.$ The
coefficient of $u_2$ give us equation
 $$u^{q+2}(2a_{{1}_{x}}+f_{u}(b_{4}-1))+u^{2q+1}qg_{1}=0$$
The equation enables us to express
$$g_{1}=\frac{1}{q}u^{1-q}(f_{u}(1-b_{4})-2a_{1_{x}})$$
 The coefficient of $u_1^2$ yields Euler equation
$$u^3(1-b_{4})f_{uuu}+u^{2}(2-qb_{4}-2b_{4})f_{uu}-uq^{2}f_{u}+q^{2}f=0.$$
 Consider for simplicity the case $b_4=1$. It is easy to see that
the last equation has two types of solutions:
 $$f= ku + nu^{-q}, \qquad q \neq -1$$
or $$f = ku + nu\ln u , \qquad q = -1 $$
 where $k, n$ are arbitrary constants.
Let us focus on $f = ku + nu^{-q} $. It follows from above
calculations and equation (9) that
$$u^{q+1}(-qa_{1_{t}}-3u^{q}qa_{1_{xx}}-4u^{q}a_{1_{xx}}+kq^{2}a_{1})u_{1}+$$
$$+nq^{2}a_{1}u_{1}+2u^{q+2}(a_{1_{tx}}-u^{q}a_{1_{xxx}}-kqa_{1_{x}})+
2una_{1_{x}}=0. \eqno (12)$$
 From (12) we have $n a_1 = 0$.

If $a_1=0$ then the solution of (9) is $h=u_2 + qu_1^2/u$. If
$n=0$ and $q \neq -4/3$ then one easily computes
 $$a_1 = (rx+s)\exp(kqt),$$
 $$ h=u_{2}+q \frac{u_1^2}{u}+
\Biggl( (rx+s)u_1 - \frac{2}{q}u r \Biggr)
 u^{-q}\exp(kqt).$$

 In the case $q = -4/3$, we obtain
 $$a_{1}=(rx^{2}+sx+p)\exp(-\frac{4}{3}kt),$$
$$h=u_{2}- \frac{4u_1^2}{3u} + \Biggl( (rx^{2}+sx+p)u_{1} +
\frac{3}{2}u(2rx+s) \Biggr) u^{4/3} \exp(-\frac{4}{3}kt).$$
 It can be shown that the found functions $h$ lead to invariant
 solutions of the corresponding equation (8).

We omit here for the sake of brevity intermediate calculations and
give the list of solutions to the equation (9):\\
 (1) if $q=-1$ and $f = su + ru\ln(u)$ then
  $$ h = u_{2} - \frac{u_1^2}{u} ; $$
 (2) if $q\neq -1$ and $f = su+ru^{-q}$ then
 $$ h = u_{2}+\frac{qu_1^2}{u} ; $$
 (3) if $q = -2$ and $f = su + ru^{3}$ then
 $$ h = u_2 - \frac{3 u_1^2}{2u} ; $$
 (4) if $q = 1$ and $f = r u $ then
 $$ h = u_2 + s\exp(rt)u^{-2}u_{1}+r/3 ; $$
 (5) if $q$ is an arbitrary constant and $f = su + ru^{1-q}$ then
 $$ h = u_{2} - \frac{(q-1)u_1^2}{u} ,$$
 with $r, s \in R.$\\
 We did not include functions $h$ that correspond to invariant
solutions of the equation (8).

If we will look for solutions to the equation (9), which depend on
third derivative, then obtain the following list:\\
 (1) if $q$ is an arbitrary constant and
 $f =su+ru^{1-q}+\frac{n(q+1)}{q^2}u^{q+1}$ then
 $$ h = u_{3}+\frac{3(q-1)}{u}
u_{1}u_{2}+(q^{2}-3q+2)\frac{u_{1}^{3}}{u^{2}}+nu_{1} ; $$
 (2a) if $q \neq 1$ and $f = nu + \frac{r}{q}u^{q+1} $ then
 $$ h = u_{3}+\frac{(3q-1)}{u}u_{1}u_{2}+
q(q-2)\frac{u_{1}^{3}}{u^2} + ru_1 ; \eqno(13) $$
 (2b) if $q = -2$ or $q= -4/3$ and $f= nu + \frac{r}{q}u^{q+1} +
 mu^{q+3}$ then $h$ is also given by (13);\\
 (3) if $q = -\frac{1}{2}$ and $f = mu$ then
 $$ h = u_3 - \frac{5u_1u_2}{2u} + \frac{5u_1^3}{4u^2}
 + r\exp(-3mt/2)u^{5/2} + s\exp(mt/2)u^{1/2} ; $$
 (4) if $q = -\frac{3}{2}$ and $f = nu + mu^{5/2}$ then
$$ h = u_3 - \frac{15u_1u_2}{2u} + \frac{35u_1^3}{4u^2}
 + r\exp(-3nt/2)u^{5/2}  ; $$
(5) if $q = -\frac{1}{2}$ and $f = mu - 2ku^{1/2}$ then
 $$ h = u_3- \frac{5u_1u_2}{2u} + \frac{5u_1^3}{4u^2} + ku_1 +
 s\exp(mt/2)u^{1/2};$$
(6) if $q = -\frac{3}{2}$ and $f = nu$ then
 $$ h = u_{3}-\frac{15u_1 u_2}{2u} +\frac{35u_1^3}
{4u^2} +s\exp(-7n/2t)u^{9/2} + r\exp(-3nt/2)u^{5/2} ; $$
 (7) if $q = -1$ and $f = mu$ then
$$ h = u_{3}-\frac{4u_1 u_2}{u} +\frac{3u_1^3} {u^2}
+s\exp(-2mt)u^2u_1  ; $$
 with $r, s, m, n \in R.$ Here we also did not include functions
 $h$ leading to invariant solutions of (8).

\vspace{15mm}
 {\bf 3. Solutions of diffusion equations.}

In this section we shall use the functions obtained above to
construct solutions of diffusion equations (8). One can apply the
method described in introduction.

We first take the function $h= u_2 + qu_1^2/u$, where $q \in R$,
corresponding to some cases mentioned above. Simply by equating
this function to zero, we obtain the differential constraint
 $$u_2 + qu_1^2/u = 0. \eqno (14)$$
The equation (14) has two types of solutions:
 $$ u = (c_1 x + c_2)^{\frac{1}{q+1}}, \qquad q \neq -1, \eqno(15)$$
 $$ u = c_1 \exp(c_2 x) , \qquad q = -1 , \eqno(16) $$
 where $c_1 , c_2$ are functions of $t$.

 If we substitute the representation (16) into equation
 $$u_t = (u_x/u)_x + ku\ln u \eqno (17)$$
then this leads us to differential equations for $c_1, c_2$. From
this equations it is easy to find $c_1, c_2$ and obtain the
following solution of (17)
 $$u = s_1\exp(s_2 x \exp(kt)) , \qquad s_1,s_2 \in R  .$$
 Substituting (15) into equation
 $$u_t = (u^q u_x)_x + su + ru^{-q} ,$$
we find the solution
 $$u =\exp(st)\Biggl( ax+b-\frac{r}{s(q+1)}\exp( -s(q+1)t )
 \Biggr)^{\frac{1}{q+1}} , $$
with $a, b \in R$.

It is easy to see that the differential constraint
 $$u_2 + (q-1)\frac{u_1^2}{u} =0 $$
for equation
 $$u_t = (u^q u_x)_x + ru + su^{1-q}  $$
 leads to solution
 $$ u = \Biggl( qax \exp(qrt) +\frac{a^2}{r}\exp(2qrt) -
 \frac{s}{r}\Biggr)^{\frac{1}{q}} , \quad   a\in R.$$

Now let us consider the differential constraints of the third
order. We start with the equation
 $$u_{t}=(u^{q}u_{x})_{x}+su+ru^{1-q}+n\frac{q+1}{q^{2}}u^{q+1} ,
\quad n\in R \eqno(18) $$
 As explained above this equation is compatible with
the differential constraint
 $$u_{3}+3(q-1)\frac{u_1u_2}{u}+(q^{2}-3q+2)
\frac{u_{1}^{3}}{u^{2}}+nu_{1} = 0. \eqno(19) $$
  By a change of variable $v = u^q$ one may rewrite (18), (19) in the
following way
 $$v_{t}=vv_{xx}+\frac{1}{q}v_{x}^{2}+n\frac{q+1}{q}v^{2} + sqv+rq ,
 \eqno (20) $$
 $$v_{3}+nv_{1} = 0. \eqno (21)$$
 If $n = -1$ then it follows from (21) that
 $$ v(t,x)= a + b\exp(x) + c\exp(-x) , \eqno(22)$$
where $a, b$ and $c$ are some functions of $t$. Substituting this
representation into equation (20) we obtain the system of ordinary
differential equations for the function $a, b$ and $c$:
 $$ a_t = -a^2(1+1/q) + a s q - 4bc/q + rq , \eqno(23)$$
 $$ b_t = - a b(1+2/q) +bsq , \eqno(24)$$
 $$ c_t = - ac(1+2/q) + csq . \eqno(25)$$
 Using finite-dimensional invariant subspaces, Galaktionov \cite{gal13}
found representation (22).

From (24) and (25) we derive the first integral $b =kc , \  k\in
R$. Therefore the system (23)-(25) can be reduced to nonlinear
ordinary differential equation for the function $a$. In general we
can not express solutions of (23)-(25) in terms of the elementary
functions. We give one example of exact solution of equation (20),
with $q=-1$ and $r=0$. This solution has the representation (22)
and the functions $a, b, c$ are
 $$a =a_1\sin(pr_1\exp(r_1 t)+m)\exp(r_1 t)/ \cos( pr_1\exp(r_1t) +m),$$
 $$b =a_2\exp(r_1t)/\cos( pr_1\exp(r_1t) +m ),$$
 $$c =a_3\exp(r_1t)/\cos( pr_1\exp(r_1t) +m) ,$$
where $a_1 = pr_1^2$, $a_3 = p^2 r_1^4/4a_2$, $r_1 = - s$ and $p,
a_2, m$ are arbitrary constants.

Now let us consider the equation
 $$u_t = (u^{-1/2}u_x)_x +mu -2k\sqrt u , \qquad m,k \in R \eqno (26)$$
and the differential constraint
 $$ u_3 - \frac{5u_1u_2}{2u} + \frac{5u_1^3}{4u^2} + ku_1
 + s e^{mt/2}\sqrt u = 0 . \eqno (27) $$
 Using the equation (26), one can write (27) as
 $$ (\ln u)_{tx} + s e^{mt/2}u^{-1} = 0 .$$
 Replacing $\ln ( e^{mt/2}u^{-1} ) $ by $w$, the last equation is
replaced by the Liouville equation
 $$ w_{tx} = s\exp(w) . $$
Since the general solution of the Liouville equation is
 $$w = \ln \frac{ 2T^{\prime}X^{\prime} }{s(T+X)^2} ,  $$
 it gives  the representation
 $$ u = \frac{s(T+X)^2}{ 2T^{\prime}X^{\prime} } e^{mt/2} , \eqno (28)$$
where $T$ and $X$ are the arbitrary functions of $t$ and $x$
respectively. Substituting this representation into (26), we have
 $$ \sqrt \frac{s}{2} \ e^{mt/4} \Biggl( 2 (T^{\prime})^{1/2} -
 (T+X)(T^{\prime})^{-3/2}T^{\prime\prime} -
 \frac{m}{2}(T+X)(T^{\prime})^{-1/2}  \Biggr) = $$
 $$ = \frac{3}{2}(X^{\prime})^{-3/2}(X^{\prime\prime})^2 -
(X^{\prime})^{-1/2}X^{\prime\prime\prime} -2k(X^{\prime})^{1/2} .
 \eqno (29) $$
 Differentiating (29) with respect to $t$, it is easy to
obtain the equation for $T$
 $$ 2T^{\prime\prime\prime}T^{\prime} - 3(T^{\prime\prime})^2 +
 \frac{m^2}{4}(T^{\prime})^2 = 0 . $$
If $m \neq 0 $ then the function
 $$ T = c_1 \tanh \Biggl( \frac{mt}{4} + c_3 \Biggr) + c_2 , \quad c_1,
 c_2, c_3 \in R $$
 is the general solution of this equation. Substituting the
function $T$ into (29), we get the following equation for $X$
 $$\sqrt{ \frac{sm}{2c_1} } \Biggl(c_1 - c_2 - X \Biggr) =
 \frac{3}{2}(X^{\prime})^{-3/2}(X^{\prime\prime})^2 -
(X^{\prime})^{-1/2}X^{\prime\prime\prime} -2k(X^{\prime})^{1/2} .
$$

 It should be noted that (28) is equivalent to the representation
 $$ u = (a_1 + a_2 e^{mt/2} )^2,    $$
 where $a_1 , a_2$ are functions of $x$. This representation yields
 the following system for $a_1$ and $a_2$
 $$  a_{ 1_{xx} } + a_1^2 m/2 - a_1 k = 0 , \eqno(30)$$
 $$  a_{2_{xx} } + a_1 a_2 m/2 - a_2 k = 0 . \eqno(31)$$
 In general, it is possible to express $a_1$ in terms of the Weierstrass
function $\wp$ and $a_2$ in terms of Lam\'{e}'s function
\cite{whi14} . However, if $ m = 12$ and $ k = 4$ , then the
functions
 $$a_1 = \frac{1}{\cosh^2(x)} , $$
 $$a_2 =  \frac{a}{\cosh^2(x)} +
  \frac{b}{\cosh^2(x)} \Biggl( \frac{\sinh 4x}{32} +
\frac{\sinh 2x}{2} + \frac{3x}{8} \Biggr) , \qquad a,b\in R $$
 satisfy the equations (30), (31).

According to our results in the previous section, as  $k=0$ , the
equation (26)  is compatible with the differential constraint
 $$  u_3 - \frac{5u_1u_2}{2u} + \frac{5u_1^3}{4u^2} +
 r e^{-3mt/2} u^{5/2}  + s e^{mt/2}\sqrt u = 0 . \eqno (32) $$
 Using the equation (26), one can write (32) as
 $$ (\ln u)_{tx} + r e^{-3mt/2}u + s e^{mt/2}u^{-1} = 0 . \eqno(33) $$
Replacing $\ln ( e^{-3mt/2}u ) $ by $w$ in (33) yields
 $$ w_{tx}  + e^w + s r e^{-w-mt} = 0 . $$
 If we set $s=0$ then from the last equation we find the following representation
 $$ u = -\frac{ 2 X^{\prime}T^{\prime} }{(X+T)^2} e^{3mt/2} , $$
where $T$ and $X$ are the arbitrary functions of $t$ and $x$
respectively. Substituting this representation into (26) leads to
equation
 $$\sqrt{-2/r} e^{3mt/4} \Biggl( -m(T^{\prime})^{1/2} (X+T) -
 2(T^{\prime})^{-1/2}T^{\prime\prime}(X+T) +4(T^{\prime})^{3/2} \Biggr) = $$
$$ =  -2X^{\prime\prime\prime}(X^{\prime})^{-3/2}(X+T)^2
+8X^{\prime\prime}(X^{\prime})^{-1/2}(X+T) - 8(X^{\prime})^{3/2} +
(X^{\prime\prime})^2 (X^{\prime})^{-5/2}(X+T)^2   $$

 Introducing new functions
$$ C(T) = \sqrt{-2/r} e^{3mt/4} \Biggl( m(T^{\prime})^{1/2} +
2T^{\prime\prime}(T^{\prime})^{-1/2} \Biggr) ,$$
 $$B(X) = (X^{\prime\prime})^2 (X^{\prime})^{-5/2}
-2X^{\prime\prime\prime}(X^{\prime})^{-3/2} , $$
 $$D(X) = 2XB + 8X^{\prime\prime}(X^{\prime})^{-1/2} , $$
 $$Q(T) = \sqrt{-2/r}e^{3mt/4} \Biggl( m(T^{\prime})^{1/2} T +
2(T^{\prime})^{-1/2}T^{\prime\prime} T - 4(T^{\prime})^{3/2}
\Biggr) , $$
 $$R(X) = BX^2 + 8X^{\prime\prime}(X^{\prime})^{-1/2}X -
 - 8(X^{\prime})^{3/2} , $$
 one can write the last equation as
 $$ C(T)X + D(X)T +B(X)T^2 + Q(T) + R(X) = 0 . \eqno (34) $$
 It is possible consider (34) as condition of orthogonality of
 two vector functions  $Z = (C,T,T^2,Q,1) , \quad W = (X,D,B,1,R) . $

Denote by $\rho(Z)$ and $\rho(W)$ the number of linearity
independent functions among $C, T, T^2, Q, 1$ and $X, D, B, R, 1$
respectively. From orthogonality condition it follows that
$\rho(Z) + \rho(W) \leq 5 .$  It is possible to show that if
$T^{\prime} \neq 0 $ then $\rho(Z) = 3$ and $\rho(W) = 2$. In this
case we have
 $$ D(X) = a_1 X + b_1 , \quad B(X) = a_2 X + b_2 , \quad R(X) = a_3 X +
 b_3 ,$$
 with $a_i , b_i \in R .$ Because of (34) and definition of the functions
$ D, B , R$ we obtain equations
 $$(X^{\prime})^3 = ( c_3 X^3 +c_2 X^2 +c_1 X +c_0 )^2
 , \eqno (35)$$
 $$(T^{\prime})^3 = A ( -c_3 T^3 +c_2 T^2 -c_1 X +c_0 )^2 , \eqno(36) $$
 where $c_3, c_2, c_1$ and $c_0$ are arbitrary constants, $A= (-2r)^{1/3}$.

 The solutions of (35) and (36) can be expressed in the terms of
the Weierstrass function $\wp$  \cite{gol15}.
 Indeed, one can
write (35) and (36) as
 $$(X^{\prime})^3 = ( c_3(X-\alpha_1)(X-\alpha_2)(X-\alpha_3) )^2
 , \eqno (38)  $$
 $$(T^{\prime})^3 = A(-c_3(T+\alpha_1)(T+\alpha_2)(T+\alpha_3)
 )^2 . $$
 Replacing $X-\alpha_1$ by $1/Y$ in (38) yields
$$(Y^{\prime})^3 + B^2(Y- b_1)^2 (Y-b_2)^2 = 0 , $$
 where $B = c_3 (\alpha_2 -\alpha_1)(\alpha_3 -\alpha_1) ,
 b_1 = \frac{1}{\alpha_2 -\alpha_1} , b_2 = \frac{1}{\alpha_3
 -\alpha_1} .$ Introducing new function $Z$ such that
 $Z^3 = B(Y- b_1)(Y-b_2)$, we obtain equation
 $$ (Z^{\prime})^2 + \frac{4B}{9}Z^3 + \frac{B^2 (b_1 -b_2)^2}{9} = 0 .\eqno(39) $$
The solutions of the last equation are expressed in the terms of
the Weierstrass function $\wp$. Applying the above process to
(36), we obtain equation such as (39).

We shall omit here other cases and discuss briefly in the next
section two-dimensional equation.

\vspace{15mm}
 {\bf 4. Two-dimensional equation. }

We consider here the fast diffusion equation
 $$ u_t = \Delta \ln (u) . \eqno(40) $$
Some applications of this equation can be found in \cite{ari16}.
Galaktionov \cite{gal10}
 used invariant subspaces to find some solutions of (40).
If we set $u = 1/v$, we obtain
 $$ v_t = v^2\Delta \ln (v) . \eqno (41) $$
It is easy to check that one of elementary solutions of (41) is travelling wave
given by
 $$ v = 1 + c\exp(mx +ny -(m^2+n^2)t) , \eqno (42) $$
where $c, m$ and $n$ are arbitrary constants. Obviously, this is invariant
solution. On the other hand, this solution satisfies differential constraints
 $$ v_x - mv +m =0 , $$
 $$ v_y - nv +n =0 . $$
It is possible to find other differential constraints that are linear with
respect $x, y$ and $v$.

It can be shown that the differential constraints
 $$v_x + \frac{c_1 - c_0 tan (t) }{c_0^2 + c_1^2}
\Biggl( v - xc_0 - yc_1 + t(c_0^2+c_1^2) \Biggr) = c_0 , $$
 $$v_y - \frac{c_0 + c_1 tan (t) }{c_0^2 + c_1^2}
\Biggl( v - xc_0 - yc_1 + t(c_0^2+c_1^2) \Biggr) = c_1   $$
 are compatible with the equation (41). Here $c_0$ and $c_1$ are
arbitrary constants. The solution of (41) corresponding to these
constraints is
 $$ v = c_0x + c_1y - (c_0^2+c_1^2)t +  c_2\cos(t) \exp(m_1x +m_2y +m_3t). $$
 Here $c_0, c_1, c_2$ are arbitrary constants and
 $$ m_1 = \frac{c_0 t - c_1}{c_0^2+c_1^2} , \quad
    m_2 = \frac{c_1 t + c_0}{c_0^2+c_1^2} , \quad
    m_3 = - tan(t) . $$
 We can derive other explicit solutions using invariant subspaces
\cite{gal10} or linear differential constraints. For example, from
\cite{gal10} one may extract the following representation
 $$ v = s_0 +s_1 \cos(x) +s_2\sin(x) +s_3\exp(y) +s_4exp(-y) , $$
where functions $s_i(t)$ satisfy ordinary differential equations
 $$s^{\prime}_{0} +s^{2}_{1} +s^{2}_{2} -4s_{3}s_{4} =0, \eqno(43)$$
 $$s^{\prime}_{1} +s_{1}s_{0}=0,$$
 $$s^{\prime}_{2} +s_{2}s_{0}=0, \eqno(44)$$
 $$s^{\prime}_{3} -s_{3}s_{0}=0, \eqno (45)$$
 $$s^{\prime}_{4} -s_{4}s_{0} = 0.$$
 Because of (44) and (45) we find
 $$s^{\prime}_{3} / s^{\prime}_{2} +s_{3} / s_{2} =0.$$
 This yields
 $$s_{2}=c_{2} / s_{3}, \quad  c_{2}\in R.$$
 By arguments similar to that used above we have
 $$s_{1}=c_{1} / s_{3}, \; s_{4}=c_{4}s_{3}, \quad  c_{1},c_{4}\in R.$$
Substituting this into (43) leads to
  $$s^{\prime}_{0} +(c_{1}^{2} +c_{2}^{2})s_{3}^{-2} -4c_{4}s^{2}_{3}=0.$$
  From (45) we express the function $s_{0}$ and obtain
  $$(\ln s_{3})^{\prime\prime} =as_{3}^{2} - bs_{3}^{-2}, \eqno(46)$$
with $a=4c_{4},\; b=c_{1}^{2} + c_{2}^{2}.$

 Setting $a=b=1$ one can derive two elementary solutions
 $$s_3 =tanh(t), \qquad   s_3 =tan(t) $$
 In general, the solutions of (46) can be expressed in terms
of elliptic functions. It is easy to obtain the correspondent
function $u$.

 The more complicated representation is
 $$ v = s_0 + s_1\cos(2x) + s_2\sin(2x) +s_3\exp(2y) +s_4\exp(-2y)
 + $$
$$ s_5\sin(x)\exp(y) +s_6\sin(x)\exp(-y) +s_7\cos(x)\exp(y)
 +s_8\cos(x)\exp(-y),$$
where $s_i$ are functions which satisfy some ordinary differential
equations. The special case of this representation was found in
\cite{gal10}.

It is important to note that the equation (40) is invariant under
infinite-dimensional algebra of symmetry \cite{dor9}. Some
solutions of (40) were obtained by means of these symmetries in
\cite{dor9}.  We shall describe other method of using symmetry. It
is convenient to apply the complex conjugate variables $z = x+iy$,
$\bar{z} = x-iy$. Thus, we can write the equation (40) as
 $$ u_t =
\frac{1}{4} \frac{\partial^2 u}{\partial z \partial \bar{z} } .
\eqno (47) $$
 It is easy to check that (47) is invariant under the
transformation
 $$ z^{\prime} = A(z) , \quad \bar{z}^{\prime} = B(\bar{z}) ,
 \quad u^{\prime} = u /(A_z B_{\bar{z} }) ,  $$
 where $A(z)$ and $B(\bar{z})$ are arbitrary functions. In other
words, if the function $f(t,z,\bar{z})$ is a solution of (47) then
$f(t,A(z),B(\bar{z}))A_z B_{\bar{z}}$ also satisfies (47).

For example, if we set $m=n=1$ then from (42) we can construct the
solution of the equation (40)
 $$ u = \frac{A_z \bar{ A_{\bar{z}}} }{1 + c\exp(A +\bar{A} - 2t) } \, ,$$
where $A$ is an arbitrary function of $z$ and $\bar{A}$ is the
complex conjugate function.

\vspace{15mm}
 {\bf 5. Conclusions. }

In sections 2 and 3 we have shown how the method of the linear
determining equations can be applied to find explicit solutions to
nonlinear diffusion equations. We have found exact solutions of
these equations, using only the simplest solutions of the linear
determining equations. It is interesting to find solutions of the
linear determining equations depending on derivatives of higher
orders. A.Shmidt \cite{shm17,shm18} applied this method to another
parabolic equations and some systems; application to the elliptic
equation is discussed in  \cite{kap8}.

In section 4 we have considered the two-dimensional equation.
Applications of systems of the linear determining equations to
multi-dimensional equations briefly discussed in \cite{kap19}.
Using results of section 3 one can find the following
representation
 $$ u = (a + b e^{mt/2} )^2 . \eqno (48)$$
 of solution of the equation
$$ u_t =  \Delta (u^{1/2}) + m u + nu^{1/2} , \qquad m,n\in R ,$$
where the functions $a(x,y)$ and $b(x,y)$ must satisfy the system
 $$ \Delta a = ma^2 + na , $$
 $$ \Delta b = m ab + nb . $$
 It is easy to show that the differential constraint
$$ u_{tt} = u_t^2 /2u + mu_t/2 $$
 leads to the representation (48). The interesting reductions of some
diffusion equations in several independent variables can be found
in \cite{gal20,rud21}. It is important to explain these reductions
by means of differential constraints.

\vspace{15mm} {\bf  Acknowledgements }

This work was supported by RFBR grant 01 - 01 - 00850, ME of
Russia grant E00 - 10 - 57 and SB of RAS grant 1.

\end{document}